\documentclass[twocolumn,showpacs,preprintnumbers,prl,aps]{revtex4}

\usepackage{graphicx}		 
\usepackage{amsmath}		
\usepackage{bm}			

\newcommand{\be}{\begin{equation}}
\newcommand{\ee}{\end{equation}}
\newcommand{\bear}{\begin{eqnarray}}
\newcommand{\eear}{\end{eqnarray}}
\newcommand{\bse}{\begin{subequations}}
\newcommand{\ese}{\end{subequations}}

\begin{document}
\title{Estimation of System Parameters in Discrete Dynamical Systems from  Time Series}	 
\author{P. Palaniyandi}
\author{M. Lakshmanan} 
\email{lakshman@cnld.bdu.ac.in}
\affiliation{Centre for Nonlinear Dynamics, Department of Physics, 
             Bharathidasan University, Tiruchirapalli - 620 024, India.} 
\date{\today}

\begin{abstract}
We propose a simple  method to estimate the parameters involved in
discrete dynamical systems from time series. The method is based
on the concept of controlling chaos by constant feedback. The major
advantages of the method are that it needs a minimal number of time
series data and is applicable to dynamical systems of any dimension.
The method also works extremely well even in the presence of noise
in the time series.  The method  is specifically illustrated by means 
of logistic and Henon maps.
\end{abstract}


\maketitle

In recent years, studies on chaotic dynamical systems have become extremely
relevant from a physical point of view due to their potential applications 
in secure communication
[1-5], cryptography [6], and so on.   Also, much attention has been given to
time series analysis since many physical, chemical, and biological systems
exhibit chaotic motion in nature.  The main objectives of time series analysis
are to identify the structure of the equations which govern the temporal
evolution of the dynamical system, the number of independent variables
involved, and parameters which control the dynamics of the system \cite{abar}. 
Several methods have been developed for modeling the dynamical systems by
different authors  \cite{e1,e2,e3,e4,e5,e6}. A number of methods have also been
proposed for estimating the system parameters based on the concept of
synchronization \cite{p1,p2,p3,p4,p5}, Bayesian approach \cite{p6,p7} 
and least squares approach \cite{p8}. In this
Letter, a very simple and practical method for estimating the system (control)
parameters of discrete dynamical systems from the time series is developed
using the concept of controlling chaos \cite{mlkm,mlsr,srml,sp}.  This method is
applicable to time series obtained from a discrete system of any dimensions
and can be extended to continuous systems also without much difficulty. The method can
also be used for the time series which contains considerable amount of noise as well
as with scalar time series.
Further this method can be used in the field of controlling chaos to find
the exact values of controlling constants ($\kappa_i$).

Consider an arbitrary $N$-dimensional discrete chaotic dynamical system
(the original map),
\bear 
x^{(n+1)}_i = f_i(x^{(n)}_1,x^{(n)}_2,...,x^{(n)}_N;{\bf p}),		
\eear 
where $i=1,2,3,...N$, ${\bf p}$ denotes the system parameters  of
dimension $M$ to be determined and the discrete index $n$ stands for denoting the
iterations.  We also assume that the function $f$ is sufficiently smooth.
Let us construct a modified discrete dynamical
system (the modified map) as
\bear 
y^{(n+1)}_i=f_i(y^{(n)}_1,y^{(n)}_2,...,y^{(n)}_N;{\bf p})+\kappa_i,
\eear
where $\kappa_i$'s are constants.  The crucial idea in the construction of the
modified map is that the addition of constants  $\kappa_i$ in Eq. (1) will not
affect the Jacobian of the original map, but it can change the original map without
affecting the parameters ({\bf p}) into  a modified map exhibiting a different stable fixed
point solution (other than the  unstable fixed point of the original map).  Also it is
always possible to construct such a  modified map  by finding a suitable set of
contants ($\kappa_i$'s) which makes the modified map to exhibit a stable
\emph{period one} fixed point even for the parameters for which the original map evolves
chaotically.  

Now let us start the evolution of the original and modified systems from a
common set of initial states  ($i.e., x^{(0)}_i=y^{(0)}_i$). After one time
interval, the dynamics  of the modified system can be represented as 
\bear
y^{(1)}_i = f_i(y^{(0)}_1,y^{(0)}_2,...,y^{(0)}_N;{\bf p})+\kappa_i   
\eear
and the dynamical variables of the original and modified  systems can be
related as
\bear
y^{(1)}_i=x^{(1)}_i+c^{(1)}_i,
\eear
where $c^{(1)}_i=\kappa_i$.  After the second interval of discrete time, the
dynamics of the modified system can expressed as
\bear
y^{(2)}_i=f_i(x^{(1)}_1+c^{(1)}_1,x^{(1)}_2+c^{(1)}_2,...,x^{(1)}_N+c^{(1)}_N;{\bf p})+\kappa_i  \; 
\eear
and, after Taylor expansion, the relation between the variable 
of the original and modified systems becomes
\bse
\bear
y^{(2)}_i=x^{(2)}_i+c^{(2)}_i,
\eear
where
\bear
c^{(2)}_i &=& \kappa_i+\sum^N_{j=1}c^{(1)}_j\frac{\partial f_i}{\partial x_j} \Big\vert_{{\bf x}^{(1)}} \nonumber \\
&&+\frac{1}{2!}			    
\sum^N_{j=1}\sum^N_{k=1}c^{(1)}_jc^{(1)}_k\frac{\partial^2 f_i}{\partial x_j\partial
x_k}\Big\vert_{{\bf x}^{(1)}}+\cdots
\eear
\ese 
and ${\bf x}$ is the vector of dimension $N$. Proceeding further, the
entire time evolution of the modified system can be obtained from the
original  system by the relation
\bse
\bear
y^{(n)}_i=x^{(n)}_i+c^{(n)}_i, \;\;\;\;\;\;\;\; n=0,1,2,...
\eear
where
\bear
c^{(n)}_i &=& \kappa_i + \sum^N_{j=1}c^{(n-1)}_j\frac{\partial f_i}{\partial x_j} \Big\vert_{{\bf x}^{(n-1)}}  \nonumber \\
 && +\frac{1}{2!}
\sum^N_{j=1}\sum^N_{k=1}c^{(n-1)}_jc^{(n-1)}_k\frac{\partial^2 f_i}{\partial x_j\partial x_k}
\Big\vert_{{\bf x}^{(n-1)}}+\cdots \;\;\;\;\;\;\;\;
\eear
\ese
and $c^{(0)}_i=0$ by our initial assumption $y^{(0)}_i=x^{(0)}_i$. 

Next, let  $ z^{(0)}_i,z^{(1)}_i,...,z^{(m-1)}_i$  be the $m$ data set  points of the given chaotic time series  
obtained for the original map. Then the trajectory of the modified map (which
is constructed by adding a set of constants $\kappa_i$ with the original map)
can be obtained from the above time series by the relation
\bse
\bear
y^{(n)}_i=z^{(n)}_i+c^{(n)}_i,
\eear
where
\bear
c^{(n)}_i  &=&  \kappa_i + \sum^N_{j=1}c^{(n-1)}_j\frac{\partial f_i}{\partial x_j} \Big\vert_{{\bf z}^{(n-1)}}  \nonumber   \\ 
   && +\frac{1}{2!}
\sum^N_{j=1}\sum^N_{k=1}c^{(n-1)}_jc^{(n-1)}_k\frac{\partial^2 f_i}{\partial x_j\partial x_k} \Big\vert_{{\bf
z}^{(n-1)}}+\cdots \;\;\;\;\;\;\;\;  
\eear 
\ese
and ${\bf z}$ is a vector  of dimension $N$.  If the original system is
one dimensional,  then
\be
c^{(n)}=\kappa +c^{(n-1)}\frac{df}{dx}\Big\vert_{z^{(n-1)}}+\frac{1}{2}
(c^{(n-1)})^2\frac{d^2f}{d^2x} \Big\vert_{z^{(n-1)}}+\cdots 
\ee 
Let $y^*_i$ be the  \emph{period one} fixed point of the modified map obtained by
Eq. (8) for the given time series data.  Then the $n^{th}$ and $(n+1)^{th}$
iterations of the map can be expressed as
\bear
z^{(n)}_i+c^{(n)}_i &=& y^*_i \;\;\;\; \text{and}	\\		 
z^{(n+1)}_i+c^{(n+1)}_i &=& y^*_i
\eear
and by subtracting Eq. (10) from Eq. (11), we get
\bse
\be
c^{(n+1)}_i -c^{(n)}_i = z^{(n)}_i-z^{(n+1)}_i.  
\ee
 Similarly, 
\be
c^{(n+2)}_i -c^{(n)}_i = z^{(n)}_i-z^{(n+2)}_i,
\ee
\ese
where $c^{(n+2)}_i$, $c^{(n+1)}_i$ and  $c^{(n)}_i$ are functions of 
$\kappa_i$ and  ${\bf p}$. Thus, we have obtained $2N$ nonlinear
simultaneous algebraic  equations  
for $(M+N)$ unknowns ($N$ $\kappa_i$'s
and $M$  ${\bf p}$'s),  and solving them we can obtain the values of the unknowns
${\bf p}$ and $\kappa_i$, provided the solution exists. 
After estimating the  unknown parameters, one
can also check the accuracy of the estimated parameters as follows.
\begin{table}[h]
\begin{center}
\caption{Convergence of r and $\kappa$ in the logistic map}   
\begin{tabular}{|c|c|c|c|c|}
\hline 
Itera- &\multicolumn{2}{|c|}{Using exact time series} & \multicolumn{2}{c|}{
Using noisy time series}  \\ \cline{2-5}  
tions &  r  & $\kappa$   & r     &  $\kappa$    \\
\hline
0 	&   10.00000000 	&  -0.50000000      	&      10.00000000 	& -0.50000000  	\\
1 	&\;  8.99885997  	&  -0.48521330   	&\;     8.99885540 	& -0.48552646		\\
2 	&\;  7.99761227  	&  -0.48347134  	&\;     7.99760776 	& -0.48374876		\\
3	&\;  6.99636547 	&  -0.48126778  	&\;     6.99636104 	& -0.48151029		\\
4	&\;  5.99512049 	&  -0.47835604  	&\;     5.99511616 	& -0.47856279		\\
5	&\;  4.99387963 	&  -0.47426330  	&\;     4.99387546 	& -0.47443116		\\
6	&\;  3.99265086 	&  -0.46786433  	&\;     3.99264699 	& -0.46798613		\\
7	&\;  3.69848466 	&  -0.46349089  	&\;     3.70025299 	& -0.46361426		\\
8	&\;  3.67047439 	&  -0.46283146  	&\;     3.67212276  	& -0.46295672		\\
9	&\;  3.67000000 	&  -0.46282092  	&\;     3.67164967 	& -0.46294621		\\
\hline 
10	&\;  3.67000000  	&  -0.46282092  	&\; 	3.67164967 	& -0.46294621		\\
\hline
\end{tabular}
\end{center}
\end{table}
Iterate the Eq. (8a)  with estimated values of the  parameters in Eq. (8b)
till  a fixed point solution ($y^*_i$) is obtained.  Then compare the fixed
point ($y^*_i$) obtained by the above iteration using the time series of
the original map with the fixed point calculated by Eq. (2) at the estimated
parameters.  The degree of closeness of these fixed points gives a measure of the accuracy
in the estimated parameters. 

As an example to our method in one dimension,   consider the well known
logistic map
\bear
x^{(n+1)} =rx^{(n)}(1-x^{(n)}),  \;\;\; 0 \leq x \leq 1, \;\;\; 0 \leq r \leq 4
\eear
where $r$ is the unknown system (control) parameter. Then the modified
logistic map can be constructed as
\bear
y^{(n+1)} =ry^{(n)}(1-y^{(n)})+\kappa,  
\eear
where $\kappa$ is a constant to be determined  which makes the
modified logistic map to exhibit  \emph{period one} fixed point solution for the
parameter   where the original map exhibits chaotic solution.  

Let $z^{(0)},z^{(1)},z^{(2)}$,...,$z^{(m-1)}$ be the  time series data obtained
from  the logistic map at some arbitrary time interval for a unknown system
parameter ($r$). Assume $z^{(0)}$ be the common initial state for both the
original and modified logistic maps ($i.e.$ $x^{(0)}=y^{(0)}=z^{(0)}$) and so
$c^{(0)}=0$ by Eq. (8a). Then after substituting three data points $z^{(1)},z^{(2)}$
and $z^{(3)}$ (one can take any three successive data) of the time series and 
the values of $c^{(1)}$, $c^{(2)}$ and $c^{(3)}$  calculated by making use of 
the Eq. (9) into the Eq. (12), we get
\bse
\bear
\kappa r(1-\kappa-2z^{(1)})&=&  z^{(1)}-z^{(2)},  \\
\kappa r(1 -2z^{(2)})[1+r(1-\kappa-2z^{(1)})] \nonumber \\
-\kappa^2 r[1+r(1-\kappa - 2z^{(1)})]^2 &=& z^{(1)}-z^{(3)}.
\;\;\;\;\;\;\;\;\;\;
\eear
\ese
The values of $\kappa$ and the unknown parameter ($r$)  can be estimated by
solving the above two nonlinear simultaneous algebraic equations  with an initial
guess of $\kappa$ and $r$.  For illustration purpose, we have used the
numerically generated time series of the logistic map for the system parameter
$r=3.67$ and solved the eqns. (15) by  globlally convergent Newton's
method \cite{nm}  with an initial guess  $-0.5$ to $\kappa$ and $10.0$ to the
parameter $r$.  The convergences of the system parameter $r$  and $\kappa$ are
shown in the Table I and it shows that the estimated value $r$ is $3.67$ which is
the exact value of parameter at which the time series data of the logistic map is
generated.   

In order to test the robustness of the method, a noisy time series  generated
by considering that the system itself produces some error in the data in each
iteration was also used in the above illustration.  In our analysis a
noise of strength $10^{-2}$ is added with the eqution of the system
\begin{widetext}
\begin{center}
\begin{table} [h]
\caption{Convergence of $\alpha$, $\beta$, $\kappa_1$ and $\kappa_2$ in the Henon map}   
\begin{tabular}{|c|c|c|c|c|c|c|c|c|}
\hline 
Itera- & \multicolumn{4}{|c|}{Using exact time series} & \multicolumn{4}{c|}{Using noisy time series}  \\ \cline{2-9}  
tions &  $\alpha$  & $\beta$   & $\kappa_1$     &  $\kappa_2$  & $\alpha$  & $\beta$   & $\kappa_1$     &  $\kappa_2$   \\
\hline
0  &   2.50000000  	&	 1.50000000 	&	-0.10000000  	&	 0.00000000 	
   & 	  2.50000000  	&	 1.50000000 	&	-0.10000000  	&	 0.00000000	\\
   
1  &   2.25643981 	&	 1.18700678 	&	-0.11800073 	&	-0.04888403		
   &	  2.25664011  	&	 1.18682541  	&	-0.11787076  	&	-0.04876757 \\
   
2  &   1.99627422 	&	 0.88915339 	&	-0.14682621 	&	-0.10147980 	
   &	  1.99662943  	&	 0.88880258  	&	-0.14654089  	&	-0.10125436	\\
   
3  &   1.72472420  	&	 0.60769535 	&	-0.19746953 	&	-0.16839182 	
   &	  1.72522662  	&	 0.60715419  	&	-0.19696823  	&	-0.16812574	\\
   
4  &   1.44784495 	&	 0.35719041	&	-0.30159310 	&	-0.26710065 	
   &	  1.44850317  	&	 0.35640699  	&	-0.30072823  	&	-0.26703903	\\
   
5  &   1.32207233 	&	 0.28601093 	&	-0.44576804 	&	-0.33566022 	
   &	  1.31986881  	&	 0.28336810  	&	-0.44645811  	&	-0.33730112	\\
   
6  &   1.41037943 	&	 0.30302200 	&	-0.45513595 	&	-0.33541800 	
   &	  1.40939982  	&	 0.30087213  	&	-0.45644206  	&	-0.33692901	\\
   
7  &   1.40001650  	&	 0.30000320 	&	-0.45914419 	&	-0.33391288 	
   &	  1.39875604  	&	 0.29768355  	&	-0.46069850  	&	-0.33545169	\\
   
8  &   1.40000000  	&	 0.30000000 	&	-0.45918942 	&	-0.33389223 	
   &	  1.39873784  	&	 0.29767999  	&	-0.46074960  	&	-0.33542922	\\
\hline   
9  &   1.40000000  	&	  0.30000000	&	-0.45918942 	&	-0.33389223 	
   &   1.39873784  	&	  0.29767999  	&	-0.46074960  	&	-0.33542922	\\
\hline
\end{tabular}
\end{table}
\end{center}
\end{widetext}
that generates the time series data. For a
particular set of data the estimated value of  $r$ is found to be $3.67164967$ after
solving eqns. (15) for the same initial guess to $\kappa$ and $r$ as before.  Also, the
values of parameter ($r$) estimated from the noisy time series data at various
intervals (we have considered 1000 data points) of time is found to be distributed around $3.67$.
We have also carried out similar analysis for the Moran-Ricker (exponential) map
and verified that the system parameter can be identified correctly both in the 
absence and presence of noise.

For the illustration of our method in two dimensional discrete system,  
we consider the Henon map,
\bse 
\bear
x^{(n+1)}_1 &=& 1+x^{(n)}_2-\alpha (x^{(n)}_1)^2,   \\
x^{(n+1)}_2 &=& \beta x^{(n)}_1, 		 
\eear
\ese
where $\alpha$ and $\beta$ are control parameters to be determined,
and the modified Henon map can be constructed as 
\bse
\bear
y^{(n+1)}_1 &=& 1+y^{(n)}_2-\alpha (y^{(n)}_1)^2 +\kappa_1,   \\
y^{(n+1)}_2 &=& \beta y^{(n)}_1 +\kappa_2,		
\eear
\ese
where $\kappa_1$ and $\kappa_2$ are constants which force the modified
Henon map to exhibit  \emph{period one} fixed point solution for a set
of parameters  where the original Henon map shows chaotic behaviour.   
 
Let $ (z^{(0)}_1,z^{(0)}_2),(z^{(1)}_1,z^{(1)}_2)$,...,$(z^{(m-1)}_1,z^{(m-1)}_2)$ 
be the data sets of time  series obtained
from the Henon map at some arbitrary interval of time for a set of unknown system
parameters ($\alpha$ and $\beta$). The starting assumption  of common initial
state $x^{(0)}_1=y^{(0)}_1=z^{(0)}_1$ and $x^{(0)}_2=y^{(0)}_2=z^{(0)}_2$ makes
$c^{(0)}_1=c^{(0)}_2=0$ by Eq. (8a). In the case of Henon map,  the substitution
of $c^{(1)}_1$, $c^{(2)}_1$, $c^{(3)}_1$, $c^{(1)}_2$, $c^{(2)}_2$ and
$c^{(3)}_2$ calculated using  Eq. (8b) into Eq. (12) leads to  
$four$ nonlinear simultaneous algebraic equations as 
\bse
\bear
	-2\alpha (\kappa_1-\alpha\kappa_1^2+\kappa_2-2\alpha\kappa_1z^{(1)}_1)z^{(2)}_1 \nonumber \\
-\alpha (\kappa_1-\alpha\kappa_1^2+\kappa_2-2\alpha\kappa_1z^{(1)}_1)^2 && \nonumber \\
+\beta\kappa_1+\kappa_2 &=& z^{(1)}_1-z^{(3)}_1, \;\;\;\;\;\;\;\;\; \\
	\beta (\kappa_1-\alpha\kappa_1^2+\kappa_2-2\alpha\kappa_1z^{(1)}_1) &=& z^{(1)}_2-z^{(3)}_2,  \\
	\kappa_2-\alpha \kappa^2_1-2\alpha\kappa_1z^{(1)}_1 &=& z^{(1)}_1-z^{(2)}_1, \\
	\beta\kappa_1 &=& z^{(1)}_2-z^{(2)}_2. 
\eear
\ese
In this illustration, we have used the numerical time series data of the Henon
map generated for the system parameters $\alpha=1.4$ and $\beta=0.3$ and solved
the above coupled Eq. (18) by the globlally convergent Newton's method
\cite{nm}  with an initial guess $\alpha=2.5$ $\beta=1.5$, $\kappa_1=-0.1$
and $\kappa_2=0$. The convergence of $\alpha$, $\beta$, $\kappa_1$ and
$\kappa_2$ are shown in Table II and it also shows that the estimated values of
$\alpha$ and $\beta$ are $1.4$ and $0.3$  respectively.  And these estimated
values are in exact agreement with the values of the parameters for which the
time series of the Henon map is generated. As in our previous example, we have
solved the Eq. (18) using the time series data containing a random noise of
strength $10^{-2}$ for the same initial guess.  In this case, the estimated
values are found to be $\alpha=1.39873784$ and $\beta=0.29767999$ and the
values of  parameters $\alpha$ and $\beta$ estimated at various interval of
time using the noisy data is found to be distributed around $1.4$ and $0.3$,
respectively.  One can also verify that the above system parameters   can be
obtained from a scalar time series (either $z_1$'s or $z_2$'s) by  contructing
four algebraic equations suitably from Eq. (12) and making use of the system
equations. For example, parameters ($\alpha$ and $\beta$) can be  estimated
from the scalar time series of $z_1$ using the four algebraic equations which
contain $z_1$ alone in the right hand side, constructed by making use of
$c^{(1)}_1$, $c^{(2)}_1$, $c^{(3)}_1$, $c^{(4)}_1$ and  $c^{(5)}_1$ in Eq.
(12).   

At this point, one may raise the question, why not invert directly the map (1)
itself using the time series data so as to find the system parameters.  While
this is certainly possible in the case of exact time series,  the extreme
sensitiveness of chaotic systems to initial conditions make it an unreliable
procedure in the presence of suitable noise. 
For example, in the case of logistic map the estimated value of $r$ is found to be 
$3.7$ while the original value is $3.67$ when an $1$\% white noise in the
range $0$ to $1$ is introduced in the time series.  On the other hand in our
method described above, no such difficulty arises.

Next, we wish to point out that it is possible to extend the analysis to 
identify the system itself in principle, say an $N^{th}$ degree polynomial  for
the right hand side of Eq. (1).  By solving sufficient number of Eqs. (12)  one
can then identify the form of the map itself.  From another point of view, the
procedure outlined here also gives a method to obtain the values of  the
controlling constants ($\kappa_i$) for a chaotic system to a desired periodic
orbit.  Finally,  we have also extended the same precedure to continuous
dynamical systems for estimating the system parameters by finding a set of
differential equations which determine the connection between the original and
modified systems. The details will be presented elsewhere.

To conclude, the main advantage of our method is that  a very minimal number  
of time series data is sufficient for the accurate  determination of the system
parameters.  We can check the accuracy of the estimated parameters  by
comparing the fixed point obtained by Eq. (8) using  the time series at
estimated parameters with  the fixed point of the modified  dynamical system at
the same parameters.  Thus, we have developed a very simple as well as  useful
method for estimating the unknown system parameters of the discrete dynamical
systems of any dimensions and illustrated it  by means of logistic and Henon
maps. 

\acknowledgements 
 
This work has been supported by the National Board of Higher Mathematics,
Department of Atomic Energy, Government of India and the Department of Science
and Technology,  Government of India through research projects.   The authors
thank Dr. K. P. N. Murthy, Dr. S. Rajasekar, Dr. K. Murali, Dr. P. Muruganandam
and Dr. P. M. Gade for many valuable suggestions.

\end{document}